\documentclass[
    ,final            
  ]
  {aipproc}

\layoutstyle{6x9}

\def\bsli{\begin{slide}}
\def\esli{\end{slide}}
\def\bitem{\begin{itemize}}
\def\eitem{\end{itemize}}
\def\be{\begin{equation}}
\def\ee{\end{equation}}
\def\bea{\begin{eqnarray}}
\def\eea{\end{eqnarray}}
\def\bear{\begin{array}}
\def\ear{\end{array}}
\def\bfig{\begin{figure}}
\def\efig{\end{figure}}
\def\bcen{\begin{center}}
\def\ecen{\end{center}}
\def\raw{\rightarrow}
\def\bra#1{\bigl\langle #1\bigr|}
\def\ket#1{\bigl| #1\bigr\rangle}

\newcommand{\matel}[3]{ \bra{#1}{#2}\ket{#3}}

\def\chic{\scriptscriptstyle}

\begin{document}

\title{Neutrino interactions with nucleons and nuclei \\
at intermediate energies}


\author{L. Alvarez-Ruso}{
  address={Institut f\"ur Theoretische Physik, Universit\"at Giessen, Germany}
}

\author{T. Leitner}{
  address={Institut f\"ur Theoretische Physik, Universit\"at Giessen, Germany}
}

\author{U. Mosel}{
  address={Institut f\"ur Theoretische Physik, Universit\"at Giessen, Germany}
}

\begin{abstract}
We investigate neutrino-nucleus collisions at intermediate energies 
incorporating  quasielastic scattering and $\Delta (1232)$ excitation as elementary 
processes, together with Fermi motion, Pauli blocking and mean-field potentials 
in the nuclear medium. A full coupled-channel treatment of  
final state interactions is achieved with a semiclassical BUU transport model. 
Results for inclusive reactions and nucleon knockout are presented.

\end{abstract}

\maketitle

The study of neutrino interactions with nucleons and  nuclei 
is crucial for current and future oscillation experiments. 
The main goal is to improve our knowledge of the fluxes, backgrounds 
and detector responses in order to minimize systematic uncertainties.
The availability of a
high intensity $\nu$ beam at Fermilab offers as well a unique opportunity to gain  
new information on the structure of the nucleon and baryonic resonances. 
Experiments such as MINER$\nu$A and FINeSSE shall address 
relevant problems like the extraction of the nucleon and $N-\Delta$ axial form factors (FF),  
or the measurement of the strange spin of the nucleon. However, those 
experiments will be performed mainly on nuclear targets. Understanding 
nuclear effects is essential for the interpretation of the data and represents both a 
challenge and an opportunity.

We have developed a theoretical model of $\nu$-nucleus collisions
at energies between 0.5 and 2~GeV, where the dominating 
processes are quasielastic scattering (QE) and $\Delta (1232)$ resonance excitation. 
The model includes all $\nu$ flavors for both charged- and neutral-current processes in any 
{\it heavy} nucleus (from $^{12}C$ on). Here we focus on the 
$\nu_\mu$ charged-current reaction on an iron target. Further details can be found in 
Ref.~\cite{Leitner:diploma}.

We describe $\nu$-nucleon interactions in a fully relativistic formalism, using state-of-the-art 
parameterizations of the FF. The hadronic current in the QE case is given by
\be
\label{QEcurrent}
\matel{p}{J^\alpha}{n} = \cos{\theta_{\chic{C}}} \, \bar u_p \left[ F_1^{\chic{V}} \left( 
\gamma^{\alpha} - \frac{q \!\!\!\!\!\!\: /\, q^{\alpha}}{q^2}  
\right) + F_2^{\chic{V}} \frac{i \sigma^{\alpha \beta} q_{\beta}}{2 m_{\chic{N}}} 
+ F_{\chic{A}} \gamma^{\alpha} \gamma_5 + F_{\chic{P}} \frac{q^{\alpha} \gamma_5}{ m_{\chic{N}}} 
\right] u_n \,.
\ee
Written in this way, the vector part of the current is conserved even if the masses 
of the initial and final nucleons differ. This is an issue in the nuclear case, where the nucleons 
have momentum and density dependent effective masses.
For the vector FF $F^{\chic{V}}_{1,2}=F^p_{1,2} - F^n_{1,2}$, we take the 
parameterization of Ref.~\cite{Budd:2003wb} (BBA-2003), which uses recent $e^-$ scattering data from JLab to 
account for deviations from the dipole $Q^2$ dependence . The axial FF is given by the standard ansatz 
$F_{\chic{A}}= g_{\chic{A}} \left(1 + Q^2/M_{\chic{A}} \right)^{-2}$, 
with $g_{\chic{A}} = -1.267$ and $M_{\chic{A}}=1$~GeV; $F_{\chic{P}}$ can be related to 
$F_{\chic{A}}$ assuming PCAC.

The $N-\Delta (1232)$ transition current involves a larger number of FF   
\be
\label{Deltacurrent}
\matel{\Delta}{J^\alpha}{N}=a \cos{\theta_{\chic{C}}} \bar \psi_\beta D^{\beta \alpha} u_{\chic{N}} \,,
\ee
with $a=\sqrt{3}$(1) for the $p-\Delta^{++}$($n-\Delta^+$) transition;   
$\psi_\beta$ is a Rarita-Schwinger spinor and 
\bea
D^{\beta \alpha } &=& \left[
  \frac{C_3^{\chic{V}}}{m_{\chic{N}}} (g^{\alpha \beta}  q \!\!\!\!\!\!\:/ - q^{\beta} \gamma^{\alpha})+
  \frac{C_4^{\chic{V}}}{m_{\chic{N}}^2} (g^{\alpha \beta} q\cdot p' - q^{\beta} p'^{\alpha}) 
+ \frac{C_5^{\chic{V}}}{m_{\chic{N}}^2} (g^{\alpha \beta} q\cdot p - q^{\beta} p^{\alpha})\right] 
\gamma_{5} \nonumber  \\  
 &&+ \frac{C_3^{\chic{A}}}{m_{\chic{N}}} (g^{\alpha \beta} q \!\!\!\!\!\!\:/ - q^{\beta} 
\gamma_{\alpha})+
  \frac{C_4^{\chic{A}}}{m_{\chic{N}}^2} (g^{\alpha \beta} q\cdot p' - q^{\beta} p'^{\alpha})+
 {C_5^{\chic{A}}} g^{\alpha \beta}
  + \frac{C_6^{\chic{A}}}{m_{\chic{N}}^2} q^{\beta} q^{\alpha}\,.
\eea
The M1 dominance of the electromagnetic $N-\Delta$ transition implies that $C_5^{\chic{V}}=0$ and 
$C_4^{\chic{V}}=-(m_{\chic{N}}/W_\Delta) C_3^{\chic{V}}$. For the remaining independent FF, we adopt 
a parameterization which describes pion electroproduction data~\cite{Paschos:2003qr}      
\be
\label{DeltaVFF}
C_3^{\chic{V}}=C_3^V(0)\left( 1+\frac{Q^2}{M_{\chic{V}}^2} \right)^{-2} 
\left( 1+\frac{Q^2}{4 M_{\chic{V}}^2} \right)^{-1}
\ee
with $C_3^{\chic{V}}(0)=1.95$ and $M_{\chic{V}}=0.84$~GeV. In the axial sector, the available 
information comes from bubble chamber experiments performed at ANL and BNL. The fits adopted the 
Adler model where $C_3^{\chic{A}}=0$ and $C_4^{\chic{A}}=-C_5^{\chic{A}}/4$. For 
$C_5^{\chic{A}}$ we have taken, as in Ref.~\cite{Paschos:2003qr}, 
\be
\label{C5A}
C_5^{\chic{A}}=C_5^A(0) \left( 1+\frac{Q^2}{M_{\chic{A}}^2} \right)^{-2} 
\left( 1+\frac{Q^2}{3 M_{\chic{A}}^2} \right)^{-1} \,
\ee
where $C_5^A(0)=1.2$ and  $M_{\chic{A}}=1.05$~GeV. Finally, $C_6^{\chic{A}}$ can be written in terms of 
$C_5^{\chic{A}}$ via PCAC as in the QE case. The correct $\Delta$ invariant mass distribution is 
implemented by means of a spectral function with an energy dependent p-wave decay width.    

When the reactions $\nu_\mu \, n \raw p \, \mu^-$ and 
$\nu_\mu \, p(n) \raw \Delta^{++}(\Delta^+) \, \mu^-$ take 
place in a nucleus, 
the initial nucleons have a finite momentum within a density dependent Fermi sea. The final 
nucleons can be Pauli blocked. This does not affect the $\Delta$'s directly, but its decay is inhibited 
due to the Pauli blocking of the nucleons that are produced in $\Delta \raw N \pi$. 
On the other side, the decay into particle-hole states causes additional broadening~\cite{Singh:1998ha}. 
We take 
also into account that nucleons and $\Delta$'s are bound in mean-field nuclear potentials 
and acquire, therefore, momentum and density dependent effective masses.     
  
\begin{figure}[h]
\label{inclusive3D}
\includegraphics[width=0.62\textwidth]{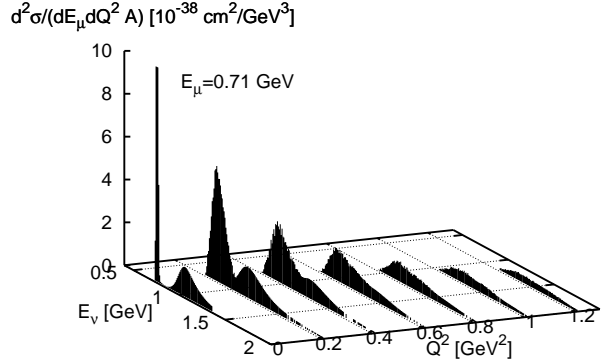}
\caption{Cross section per nucleon for the inclusive process $\nu_\mu \, ^{56}Fe \raw X \, \mu^-$.}
\end{figure}

The double-differential cross section for the inclusive reaction $\nu_\mu \, ^{56}Fe \raw X \, \mu^-$ 
is shown in  Fig.~\ref{inclusive3D} as a function of the neutrino energy and the 4-momentum transfer 
at a fixed energy of the outgoing muon. At low $Q^2$ one can clearly distinguish two peaks associated 
with QE scattering (at lower energies) and $\Delta$ excitation. As $Q^2$ increases, the peaks 
get broader and overlap, while the cross section tends to zero. 

In order to describe exclusive channels, final state 
interactions (FSI)  have to be managed. This is achieved with a semiclassical BUU transport model in 
coupled channels (cf.~\cite{Effenberger:1999ay}). The produced particles, moving 
along classical trajectories in the mean-field nuclear potential, can undergo elastic and inelastic 
collisions with the nucleons, decay ($\Delta$'s) or be absorbed (pions).

\begin{figure}[h]
\label{knockout}
\begin{minipage}{0.47\textwidth}
\includegraphics[width=\textwidth]{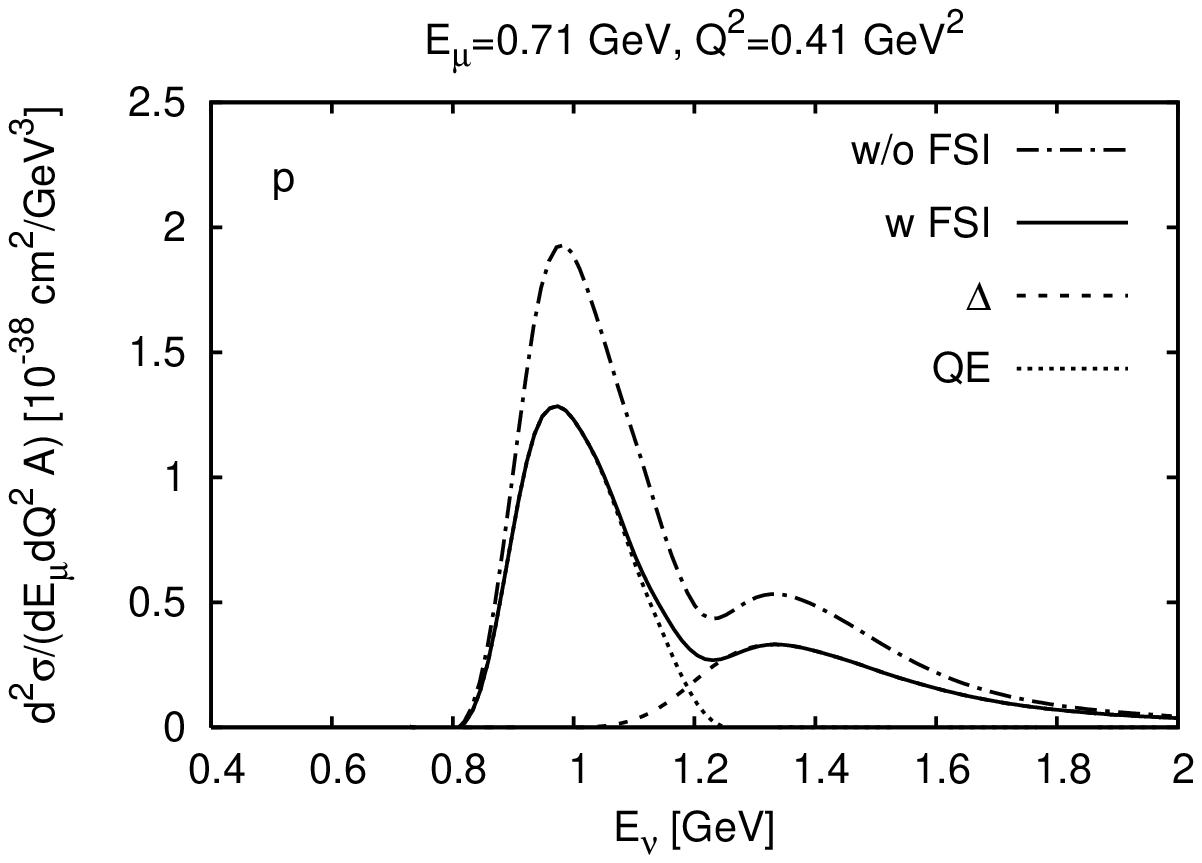}
\end{minipage} \hfill
\begin{minipage}{0.47\textwidth}
\includegraphics[width=\textwidth]{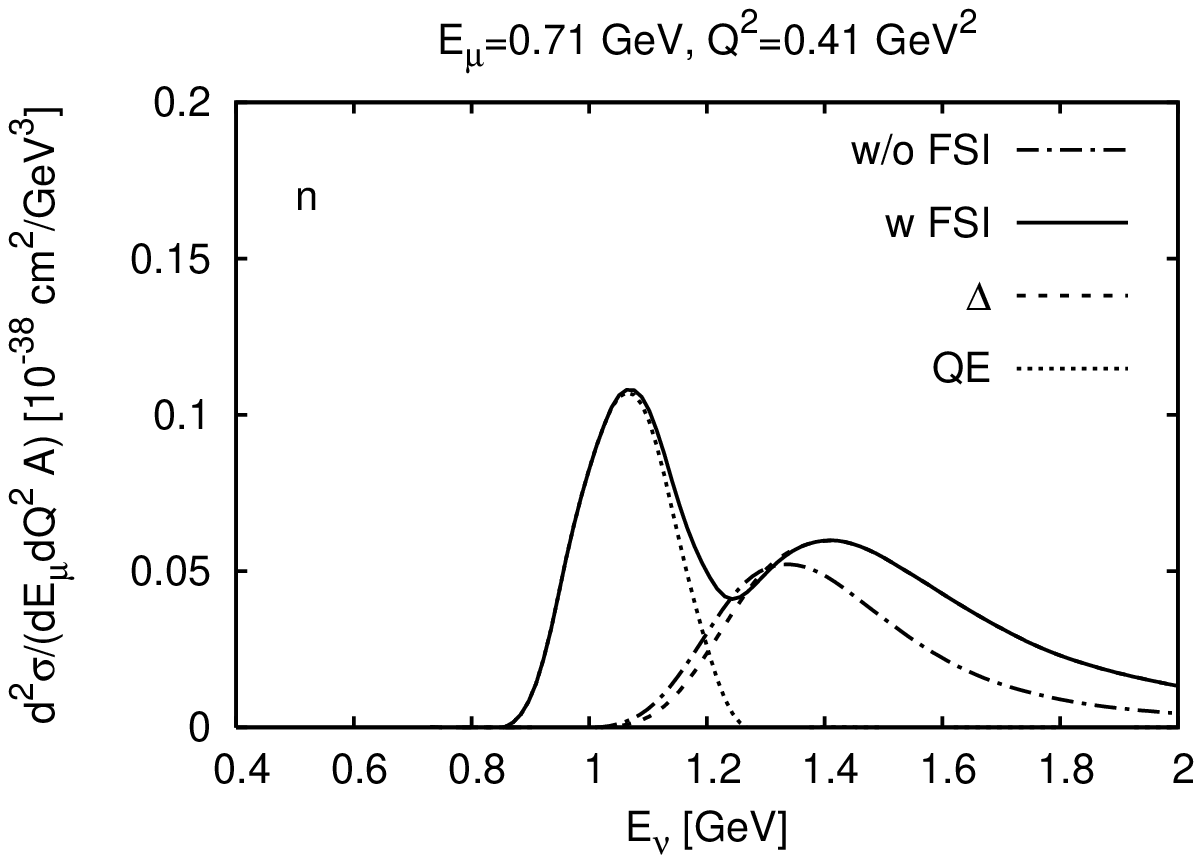}
\end{minipage} 
\caption{Proton and neutron knockout cross sections in charged-current $\nu_\mu \, ^{56}Fe$ interactions.}
 \end{figure}

A first sample of our results for $\pi$ production can be found in Ref.~\cite{Leitner:2005jg}. 
In Fig.~\ref{knockout} we present our distributions for 
$p$ and $n$ knockout as a function of the $\nu$ energy for fixed values of $E_\mu$ and $Q^2$.
The overlaping QE and $\Delta$ peaks are clearly visible. As for QE, only
$p$'s can be produced in the initial collision; $n$'s emerge only as a result of FSI 
processes. This explains the small cross section for $n$ knockout and the reduction of the $p$ one when 
FSI is considered. In the $\Delta$ region the situation is similar since only a small fraction of 
the neutrons comes directly from $\nu_\mu \, n \raw \Delta^+ \, \mu^- \raw n \, \pi^+ \,\mu^-$.   

Work supported by the Deutsche Forschungsgemeinschaft.


\bibliographystyle{aipproc.bst}
\bibliography{biblioneutrino}

\end{document}